\documentclass[letter]{jpsj2} %% for letters
\usepackage{color}

\title{
Electronic Structure of Superconducting FeSe Studied by High-Resolution Photoemission Spectroscopy
}

\author{Rikiya \textsc{Yoshida}$^{1}$%%\thanks{Multiple authors and affiliations correspond using arabic numerals each other.}
, Takanori \textsc{Wakita}$^{2, 3}$
, Hiroyuki \textsc{Okazaki}$^{1}$
, Yoshikazu \textsc{Mizuguchi}$^{4, 5}$
, Shunsuke \textsc{Tsuda}$^{3, 4, 6}$
, Yoshihiko \textsc{Takano}$^{3, 4, 5}$
, Hiroyuki \textsc{Takeya}$^{4}$
, Kazuto \textsc{Hirata}$^{4}$
, Takayuki \textsc{Muro}$^{7}$
, Mario \textsc{Okawa}$^{8}$
, Kyoko \textsc{Ishizaka}$^{8}$
, Shik \textsc{Shin}$^{8, 9}$
, Hisatomo \textsc{Harima}$^{3, 10}$
, Masaaki \textsc{Hirai}$^{1, 2, 3}$
, Yuji \textsc{Muraoka}$^{1, 2, 3}$%%\thanks{E-mail: xxxxx@vvvv.com}
, and Takayoshi \textsc{Yokoya}$^{1, 2, 3}$%%\thanks{Present address:  Department of Applied Physics, University of ABCDE, Tokyo.}
}

\inst{
$^{1}$The Graduate School of Natural Science and Technology, Okayama University, Okayama 700-8530, Japan \\
$^{2}$Research Laboratory for Surface Science (RLSS), Okayama University, Okayama 700-8530, Japan \\
$^{3}$JST, Transormative Research-Project on Iron Pnictides (TRIP), Chiyada, Tokyo 102-0075, Japan \\
$^{4}$National Institute for Material Science, Tsukuba, Ibaraki 305-0047, Japan \\
$^{5}$University of Tsukuba,  Tsukuba, Ibaraki 305-8577, Japan \\
$^{6}$WPI-MANA-NIMS, Tsukuba, Ibaraki 305-0044, Japan \\
$^{7}$Japan Synchrotron Radiation Research Institute (JASRI)/SPring-8, Sayo, Hyogo 679-5198, Japan \\
$^{8}$Institute for Solid State Physics, The University of Tokyo, Kashiwa, Chiba 277-8581, Japan \\
$^{9}$The Institute of Physical and Chemical Research (RIKEN), Sayo, Hyogo 679-5143, Japan \\
$^{10}$Department of Physics, Kobe University, Kobe 657-8501, Japan \\
}

\abst{We have performed soft x-ray and ultrahigh-resolution laser-excited photoemission measurements on tetragonal FeSe, which was recently identified as a superconductor. Energy dependent study of valence band is compared to band structure calculations and yields a reasonable assignment of partial densities of states. However, the sharp peak near the Fermi level slightly deviates from the calculated energy position, giving rise to the necessity of self-energy correction. We have also performed ultrahigh-resolution laser photoemission experiment on FeSe and observed the suppression of intensity around the Fermi level upon cooling.}

\kword{iron-based superconductor, FeSe, electronic structure, photoemission spectroscopy}

\begin{document}
\maketitle

%%\section{Introduction} %% No sections necessary for express letters, letters and short notes
%%\par
Triggered by the discovery of high temperature superconductivity in the system of LaFeAsO$_{1-x}$F${_x}$ (known as `1111 system'),~\cite{Kamihara} the field of iron-based superconductors has been growing at enormous rate.~\cite{Takahashi, Cruz, Rotter_PRB, Ishida, Sato, Ding, Kondo} While the connection between this iron-based superconductor and high temperature cuprates was the first interest of people, discovery of new types such as (Ba$_{1-x}$K$_{x}$)Fe$_{2}$As$_{2}$ (`122 system')~\cite{Rotter_PRL} and Li$_{1-x}$FeAs (`111 system'),~\cite{Wang, Tapp} opens up a new way for our approach, that is, the comparison among iron-based superconductors. It is probably safe to say that the comparison among the iron-based system should give us a lot of information on their nature of superconductivity.
\par
On the basis of current research status mentioned above, it is important to mention the possibility to take another approach presented by Hsu {\it et al}.~\cite{Hsu} Very recently, Hsu {\it et al}. reported the occurrence of superconductivity in tetragonal FeSe (or $\alpha$-FeSe) at 8 K.~\cite{Hsu} The crystal structure of tetragonal FeSe (space group: {\it P}4/nmm) is actually the simplest among the recently found iron-based superconductors because it is only composed of stacking tetrahedrally-bounded FeSe$_{4}$ layers. FeSe is quite appealing due to this structural simplicity and the relative easiness of theoretical handling, but studies on the nature of FeSe have just started. For instance, a study revealed that critical temperature of FeSe increases under high pressure, yielding the onset of 27 K at 1.48 GPa.~\cite{Mizuguchi} While FeSe goes through structural phase transition around 70 K,~\cite{Margadonna} the result of NMR showed the nuclear-spin lattice relaxation rate 1/$T_{1}$ follows $T^{3}$ behavior below the critical temperature ($T_\mathrm{c}$), suggesting an unconventional superconductivity.~\cite{Kotegawa} Moreover, A. Subedi {\it et al}. have performed density functional study on the electronic structures of FeSe, reporting that the density of states (DOS) at the Fermi level ($E_\mathrm{F}$) is dominated by Fe {\it d}-state.~\cite{Subedi} While the last theoretical study provides a basic information on the electronic structure of FeSe, there is no corresponding experimental report on tetragonal FeSe to the best of our knowledge. Therefore, the importance to investigate the electronic structures of FeSe using photoemission spectroscopy must be extremely high.
\par
Here we report our experimental results of soft x-ray and ultrahigh-resolution laser-excited photoemission spectroscopies on FeSe to provide a basic information of its electronic structure. We compared the result of valence band measurements to the first principle calculation and discuss the assignment of partial densities of states. We also measured the near $E_\mathrm{F}$ spectra in ultrahigh-resolution and observed that the intensity around the Fermi level becomes suppressed upon cooling. Throughout this paper, comparisons between FeSe and other iron-based superconductors are commented.
\par
We prepared polycrystalline samples of FeSe employing the standard method of solid state reaction. The detail of the sample synthesis is described elsewhere.~\cite{Mizuguchi} We characterized these samples with x-ray diffraction method, observing the small peaks of hexagonal FeSe as an impurity phase. Magnetization measurements on our samples confirmed a sharp superconducting transition at 8 K. Note that, as previously reported, our FeSe samples are slightly deficient in selenium (FeSe$_{1-x}$, where  x$\sim$0.08).~\cite{Margadonna}
\par
Our soft x-ray photoemission measurements were performed at two different sites. For the experiment at BL-5 (Okayama University Beamline) of Hiroshima Synchrotron Radiation Center (HiSOR), monochromatized x-rays of 140 eV were employed for photoelectron excitations with total energy resolution of 200 meV, and measurements were done at room temperature with the base pressure better than $5\times 10^{-10}$ Torr. On the other hand, the experiment at BL27SU of SPring-8 used photons of 1100 eV with total energy resolution of 180 meV. In this latter experiment, a PHOIBOS 150 electron analyzer was employed, and measurements were done at 13 K with pressure better than $5.5\times10^{-10}$ Torr. Binding energies were calibrated in reference to the Fermi edge of molybdenum, which was electrically linked with the sample, or the Fermi edge of gold thin film evapolated near the sample. For surface preparation, we fractured or scraped the samples {\it in situ} in both experiments.
\par
In order to study the electronic structure near the Fermi level, we have employed ultrahigh-resolution laser-excited photoemission spectroscopy (laser-PES). The system consists of a vacuum ultraviolet laser (photon energy of 6.994 eV) and a Gammadata-Scienta R4000WAL electron analyzer, and extremely bulk-sensitive measurements are possible using this system.~\cite{LaserPES} Samples were fractured {\it in situ}, and photoemission spectra were taken in the temperature range of 6 - 25 K. Binding energies of the spectra were calibrated in reference to the Fermi edge of a gold film evaporated near the sample, and the total energy resolution should be better than 6 meV. The base pressure of main chamber was kept better than $2\times10^{-11}$ Torr through the experiment. 
\par
For comparison, we have carried out band structure calculations using the full-potential augmented plane wave method (FLAPW) with the local density approximation (LDA). Following the LDA formalism proposed by Gunnarsson and Lundqvist,~\cite{GL} densities of states were calculated with the program package codes KANSAI-04 and TSPACE on the scalar relativistic scheme with the spin-orbit interaction included as a second variational procedure. For crystallographic parameters, we employed the values experimentally determined on the low-temperature phase.~\cite{Margadonna} Even though our FeSe samples are slightly deficient in selenium, $E_\mathrm{F}$ can be higher no more than 20 meV. Therefore, it would not affect the discussion given in this report.
\par
Figure 1 shows the valence band photoemission spectra of tetragonal FeSe taken with two different photon energies. The measured spectra have 4 distinct features: a sharp peak near $E_\mathrm{F}$ (feature {\bf A}), another peak at 2.0 eV (feature {\bf B}), and two peaks at around 4.0 eV and around 5.8 eV (feature {\bf C} and {\bf D}), where feature {\bf C} and {\bf D} are more apparent in the 1100 eV spectrum. The difference in the spectral shape should come from the cross-section effect, where the cross-sectional ratio of Fe {\it d}-state to Se {\it p}-state is the order of 10$^{2}$ for the 140 eV photons but is the  order of unity for 1100 eV photons.~\cite{CS} We expect that the spectrum taken with 140 eV photons reflects Fe {\it d}-state dominantly, and the one with 1100 eV photons should do both Fe {\it d}-state and Se {\it p}-state. Therefore, from the observation, it is suggested that Fe {\it d}-state dominates the density of states at $E_\mathrm{F}$.
\par
We display the result of the band structure calculation in Fig. 2. The calculation shows that Fe {\it d}-state dominates the states near $E_\mathrm{F}$, and Se {\it p}-state mainly lies around 3 - 6 eV region. Note that our result is consistent with the result reported by A. Subedi {\it et al}.~\cite{Subedi} We can compare the calculated partial densities of states to the observed features as follow: feature {\bf A} and {\bf B} should correspond to Fe {\it d}-state; feature {\bf C} shows hybridized state of Fe {\it d}-state and Se {\it p}-state; and feature {\bf D} is mainly Se {\it p}-state. These assignments are reasonably supported by the fact that while feature {\bf A} and {\bf B} remain prominent in both spectra, the intensities of feature {\bf C} and {\bf D} are enhanced in the 1100 eV spectrum. Since the dominance of Fe {\it d}-state is also experimentally supported for other iron-based superconductors, ~\cite{Sato} it should be attributed to an important similarity of FeSe with them. 
\par
The inset of Fig. 2 shows the expanded near $E_\mathrm{F}$ spectrum taken with 1100 eV photons and the calculated DOS. Note that the peak at around the binding energy of 0.3 eV corresponds to the sharp peak (feature {\bf A}) appeared in Figure 1. One can notice that the peak position deviates from the highest peak in the calculated total DOS, approximately by 0.25 eV. Unfortunately, our measurement with 1100 eV photons does not allow sufficient resolution for self-energy analysis; however, interestingly, such deviation was also observed for LaFeAsO and LaFePO.\cite{Malaeb} Malaeb {\it et al}. claim the necessity of self-energy correction to the band calculations and estimate the mass enhancement $m^{*}/m_{b}$ ($m^{*}$: enhanced mass at $E_\mathrm{F}$, $m_{b}$: bare band mass) to be $\simeq$ 1.8 for LaFeAsO$_{1-x}$F$_{x}$ and $\simeq$ 1.5 for LaFePO$_{0.94}$F$_{0.06}$. It is noted that similar discussion of self-energy correction is also given in the experimental study of  a ferromagnetic metal F$_7$Se$_8$, and $m^{*}/m_{b}$ $\simeq$ 1.7 is reported for this compound.~\cite{Shimada}  Qualitatively, the deviated peak position in FeSe spectrum is similar to what is reported for these compounds, thus measurements with higher energy resolution could facilitate the understanding of the correlation effects in superconducting FeSe and possibly give the value of $m^{*}/m_{b}$ in the same order.
\par
Ultrahigh-resolution near $E_\mathrm{F}$ spectra of FeSe, taken with the laser-excited photoemission spectrometer in the temperature range of 6 - 25 K, is shown in Fig. 3. Observed spectra seem to emerge not at $E_\mathrm{F}$ but off $E_\mathrm{F}$ (see the inset), and the intensity around $E_\mathrm{F}$ tends to decrease upon cooling. The spectra measured at 6 K should correspond to the superconducting state, but no clear quasi-particle peak was observed. While it is hard to conclude that our observation captures the opening of superconducting gap, similar spectra with no clear quasi-particle peak have been reported for the system of LaFeAsO$_{1-x}$F${_x}$ below $T_\mathrm{c}$.~\cite{Ishida, Sato} In order to elaborate our analysis, we symmetrized the spectra as illustrated in the same figure. 
In the symmetrized spectra, we clearly see the reduction of intensity around $E_\mathrm{F}$ upon cooling. Suppose this suppression of intensity below $T_\mathrm{c}$ indicates the opening of superconducting gap,  we can speculate that the broad spectrum is the indication of unconventional pairing in FeSe, as previously suggested by nuclear magnetic resonance experiment.~\cite{Kotegawa} However, extreme care is indispensable to conclude the origin of this intensity suppression at $E_\mathrm{F}$, so further studies are required on this compound. It is noted that even though angle-resolved photoemission spectroscopies (ARPES) on Ba$_{0.6}$K$_{0.4}$Fe$_{2}$As$_{2}$ and NdFeAsO$_{0.9}$F$_{0.1}$ have shown the realization of nodeless pairing in them,~\cite{Ding, Kondo} it is still an open question whether FeSe share the same symmetry. Therefore, momentum-resolved measurements on single crystalline FeSe is also encouraged.
\par
In conclusion, we have presented the results of soft x-ray and ultrahigh-resolution laser-excited photoemission measurements on tetragonal FeSe. Observed features in the valence band spectra were reasonably assigned by the band calculation, confirming the dominance of Fe {\it d}-state at $E_\mathrm{F}$. The deviation between experimental spectrum and band calculation implies that self-energy correction is necassary to describe FeSe. Temperature dependence of ultrahigh-resolution near $E_\mathrm{F}$ spectra shows the suppression of intensity at $E_\mathrm{F}$ upon cooling. These behaviors are similar to what is observed for other iron-based superconductors, suggesting the similarity of electronic structures between FeSe and other iron-based superconductors. To understand the observed suppression of intensity at the Fermi level, further studies are necessary. Moreover, comparison between FeSe and high $T_\mathrm{c}$ would be helpful to understand the nature of this interesting superconductor.
\par
We would like to thank Prof. M. Ichioka for fruitful discussions. We also thank Dr. T. Kiss and Mr. Y. Nakamura for their help on experiments. For experiments at HiSOR, we acknowledge Prof. M. taniguchi, Prof. H. Namatame, and other stuffs there. The synchrotron photoemission measurements at SPring-8 were done under the proposal number of 2008B1581. This work is supported by Grant-in-Aid for Scientific Research from the Ministry of Education, Culture, Sports, Science and Technology.

\begin{figure}[t]
\begin{center}
\includegraphics[width=204pt]{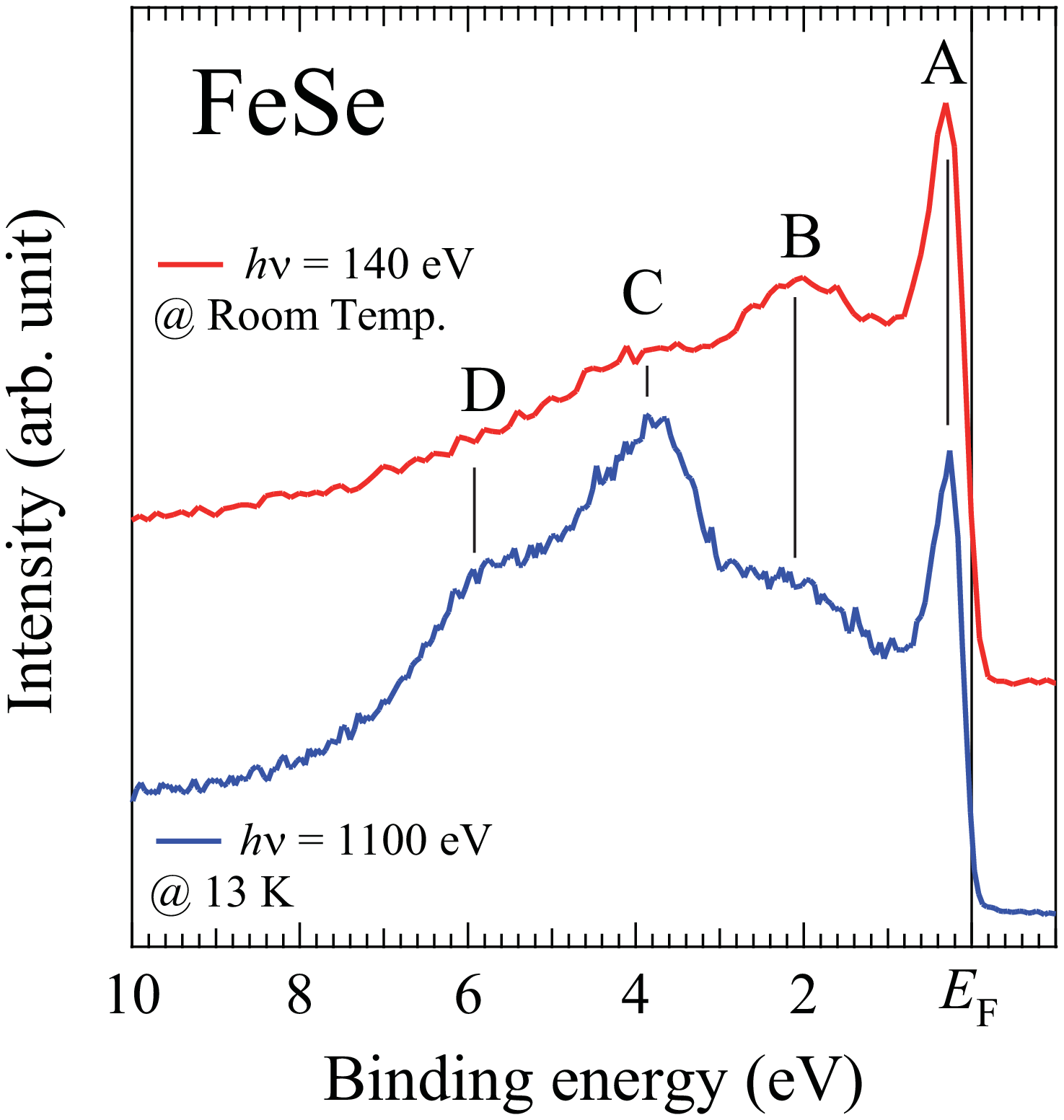}
\end{center}
\caption{(Color online) Valance band spectrum of FeSe measured at room temperature with photon energy of 140 eV and the one measured at 13 K with soft x-rays of 1100 eV.}
\label{Fig1}
\end{figure}
\begin{figure}[t]
\par
\begin{center}
\includegraphics[width=204pt]{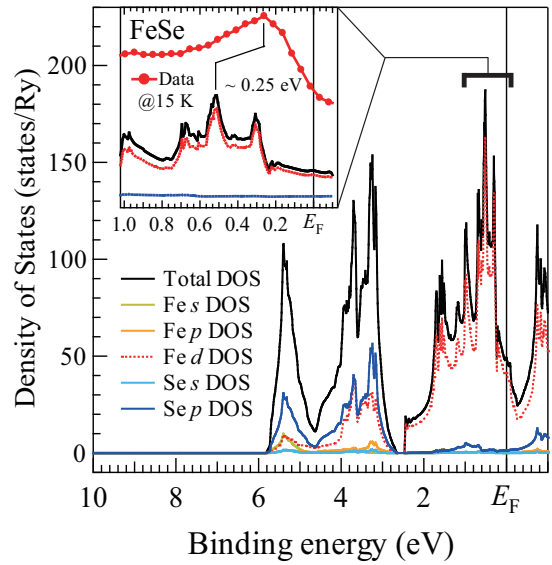}
\end{center}
\caption{(Color online) Total and partial densities of states calculated for FeSe. The inset compares the experimental data taken with 1100eV photons with the calculation near the Fermi level.}
\label{Fig2}
\end{figure}
\par
\begin{figure}[t]
\begin{center}
\includegraphics[width=245pt]{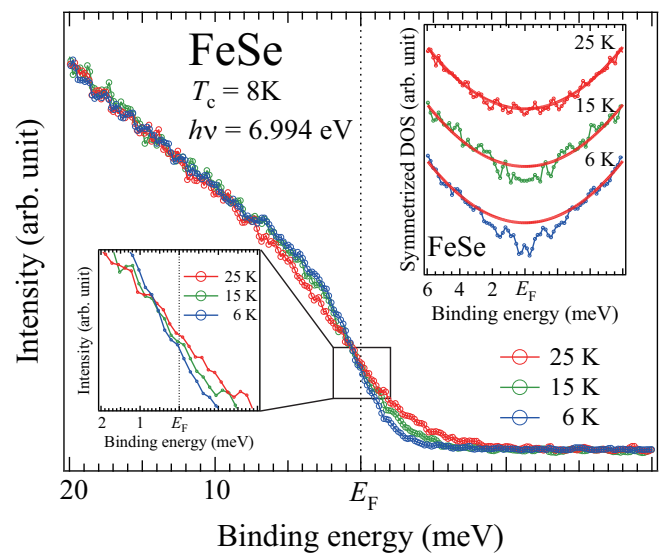}
\end{center}
\caption{(Color online) Ultrahigh-resolution spectra near $E_\mathrm{F}$ of FeSe measured with photon energy of 6.994eV in the temperature range of 6 - 25 K.}
\label{Fig3}
\end{figure}
\end{document}